\documentclass[epj]{svjour}
\usepackage{graphics}
\begin{document}
\title{Long-range chiral dynamics of $\Lambda$-hyperon in nuclear media}
\author{J. Mart\'in C\'amalich \thanks {This work was partially supported by DGI and FEDER funds,
contract  BFM2003-00856 and  by the EU Integrated Infrastructure
Initiative Hadron Physics Project contract RII3-CT-2004-506078. The author
acknowledges support from  CSIC-Fundaci\'on Bancaja.} }                  
\institute{
\email{camalich@ific.uv.es }\\
Departamento de F\'isica Te\'orica e IFIC, Universidad de Valencia-CSIC;\\
Institutos de Investigaci\'on de Paterna, Aptado. 22085, 46071 Valencia, Spain}
\date{Received: date / Revised version: date}
%
\abstract{
We extend a chiral effective field theory approach to the $\Lambda$-nuclei interaction 
 with  the inclusion of the decuplet baryons. More precisely, we 
study the contributions due to the long-range two-pion exchange,
with $\Sigma$ and $\Sigma^*$ baryons in the internal baryonic lines considering
Nh and $\Delta$h excitations.  In particular, central and spin-orbit potentials are
studied. For the former, regularization is needed and physical values of the 
cut-off give a large attraction, becoming
necessary to include the repulsion of other terms not considered here. For the latter, in a model-independent framework, the inclusion of the
decuplet supports the natural 
explanation of the smallness of the $\Lambda$-nuclear spin-orbit term 
and shows the importance of the $\Sigma^*$ and $\Delta$ degrees of  freedom for 
the hyperon-nucleus interactions.
\PACS{
      {21.80}{Hypernuclei}   \and
      {21.65.+f}{Nuclear matter} \and
      {13.75.Ev}{Hyperon-nucleon interaction}\and
      {24.10.Cn}{Many-body theory}
     } 
} 
\maketitle
\section{Introduction and framework}
\label{intro}
One of the most striking features of the $\Lambda$-nucleus potential is the 
weakness of the spin-orbit piece \cite{Bouyssy:1977jj,Bruckner:1978ix}. Several
theoretical approaches have tried to explain it, ranging
from one boson exchange (OBE) potentials \cite{Noble:1980kg} with the couplings sometimes motivated by the
underlying quark dynamics, to the consideration of two meson exchange pieces 
\cite{Brockmann:1977es} or to quark based models
\cite{Pirner:1979mb}. Recently, a novel approach to the 
$\Lambda$-nucleus interaction based on effective field theory has attempted to
explain this fact as a natural cancellation of short and long-range pieces \cite{Kaiser:2004fe}. The
main contribution to the $\Lambda$ mean field in that work comes from the first diagram
in Fig.1, which is related to the nucleon-hole excitation contribution to the pion
self-energy. However, we know from pion physics the importance of $\Delta$-hole excitation
even at very low energy well below the $\Delta$ peak \cite{Oset:1981ih}. Also, in Refs.
\cite{Holzenkamp:1989tq,Sasaki:2006cx}  it was shown the importance of the
decuplet baryons as intermediate states in the two meson exchange terms of the
$YN$ bare potential. The aim of this work is 
to extend the Ref. \cite{Kaiser:2004fe} considering also the interaction
with the  relevant baryons of the decuplet ($\Delta$ and $\Sigma^*$) and its 
contribution to the two pion exchange potential. In particular, we will study  whether the natural explanation of
the weakness of  the spin-orbit $\Lambda$-nucleus potential is still valid after the
inclusion of the new terms. This paper reports briefly
the results obtained in \cite{Camalich:2006is}.

The coupling between the pseudoscalar meson octet and the baryon octet is
given by
\begin{eqnarray}
{\cal L_{\rm oct}} = D\,{\rm Tr}\big( \bar{B} \gamma_\mu\gamma_5\{u^\mu,B\}\big)+F\,
{\rm Tr}\big(\bar{B} \gamma_\mu\gamma_5\big[u^\mu,B\big]\big), \label{eq:lagoctet}
\end{eqnarray} 
where $B$ is the traceless flavour matrix accounting for the spinors fields of baryons
while $u^\mu=i[\xi^\dagger,\partial^\mu\xi]/2$ introduce the SU(3) matrix
of meson fields $\Phi$ since $\xi=Exp(i \Phi/\sqrt{2}f_\pi)$. The interaction between the
baryon octet, the baryon decuplet and the meson octet is described by \cite{Butler:1992pn}:

\begin{figure*}
\begin{center}
\resizebox{0.75\textwidth}{!}{%
\includegraphics{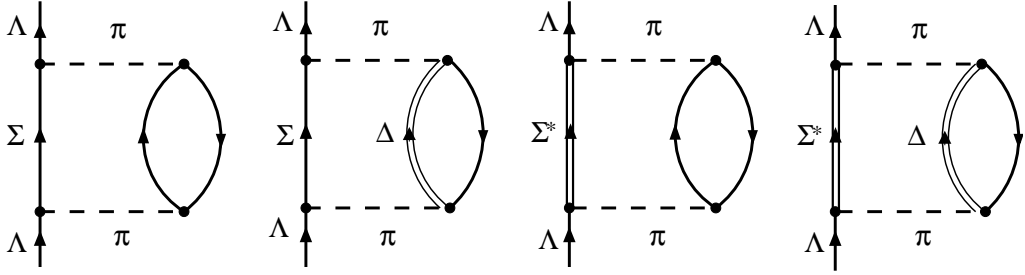}}       
\caption{Two pion exchange diagrams with $\Sigma$, $\Sigma^*$ in intermediate states and with
1Nh and 1$\Delta$h excitations of nuclear matter. Nucleon in-medium propagators are used.}      
\end{center}
\end{figure*}
\begin{eqnarray}
{\cal L_{\rm dec}} = \frac{{\cal C}}{\sqrt{2}f_\pi}\big(\bar{T_\mu} \partial^\mu \phi B + 
{\rm h.c.}\big),   \label{eq:lagdec}
\end{eqnarray}
where T$^\mu_{abc}$ is the SU(3) representation for the 3/2$^+$ spinor fields and
where we have expanded the axial current up to one meson field.
The parameter $f_\pi$=92.4 MeV is the weak pion decay constant and D=0.84, F=0.46 are the SU(3) 
axial-vector coupling constants for the octet baryons. The analysis of the partial decay widths of the decuplet shows a breaking of the SU(3) symmetry
~\cite{Butler:1992pn,Doring:2006ub} of the order of 30\%. In our calculation we need the
$\Sigma^*\pi\Lambda$ and $\Delta\pi N$ vertices and we will use for each case
as coupling constant ${\cal C}$ the value fitted to the decay widths of  $\Sigma^*\rightarrow\pi\Lambda$  
and $\Delta\rightarrow\pi N$ respectively (${\cal C}_{\Sigma *} =1.7,\,{\cal C}_{\Delta }=2.0$).

\section{$\Lambda$-nucleus central potential depth}
\label{sec:1}

We focus on the obtainment of the density dependence of the mean-field $U_\Lambda(k_f)$ for a
zero-momentum $\Lambda$-hyperon interacting with an infinite isospin-symmetric nuclear matter
via the long-range terms represented in Fig.1, (OPE interaction is prohibited by isospin 
conservation). The nucleon lines represent in medium nucleon propagators
\begin{eqnarray}
	G(p)=\frac{\theta(|\vec{p}|-k_f)}{\gamma\cdot
	p-m+i\epsilon}+\frac{\theta(k_f-|\vec{p}|)}{\gamma\cdot p-m-i\epsilon}\,,
\end{eqnarray}
where $k_f$ is the Fermi momentum. We start with the first diagram of Fig.1 reproducing the
procedure of \cite{Kaiser:2004fe}. Two pieces, direct and crossed, appear in the pion 
self-energy after the N-hole loop energy integration. Moreover, each of them can be split into
two other pieces, linear and quadratic in the occupation number $n(k)=\theta(k_f-|\vec{k}|)$.      

A non-relativistic reduction
is performed expanding  the potentials in power series of a variable
representative of the average mass of baryons, $M_B\equiv(2\,M_N+M_\Lambda+M_\Sigma)/4$. Other
relevant magnitudes are the mass splittings either on the internal-baryon line or on the 1ph excitation,
expressed in a convenient way;  $M_\Sigma-M_\Lambda\equiv\Delta^2/M_B\simeq 77$ MeV, $M_{\Sigma^*}-M_\Lambda\equiv {\Delta^*}^2/M_B \simeq
292$ MeV for the internal-line and $M_\Delta-M_N\equiv\Delta_f^2/M_B\simeq 270$ MeV for the 1ph. 

Once integrated the energy in both loops, the direct term proportional to occupation number in the 
first diagram of Fig.1 gives to leading order the following contribution to the mean-field:
\begin{eqnarray}
{\small U_\Lambda(k_f)^{\rm Nh-l}=-\frac{D^2 g_A^2}{f_\pi^4}\int_{\mid\vec{p_1}\mid
<k_f}\frac{d^3p_1d^3l}{(2\pi)^6}\times} \nonumber \\
{\small \times \frac{M_B\,\vec{l}^4}{(m_\pi^2+\vec{l}^2)^2(\Delta^2+
\vec{l}^2-\vec{l}\cdot\vec{p_1})},} \label{eq:dirlin}
\end{eqnarray} 
where $g_A=D+F=1.3$, $m_\pi$=138 MeV, $\vec{l}$ is the trimomentum of the 
pion and $\vec{p_1}$ is the trimomentum of the nucleon. The $d^3l$ loop is linearly ultraviolet divergent and we 
regularize it with a cut-off $\bar{\Lambda}=0.6$ GeV. 
The direct term, quadratic in occupation number in Fig.1 is a repulsive piece that, at leading order, 
gives
\begin{eqnarray}
{\small U_\Lambda (k_f) ^{\rm Nh-q} = \frac{D^2 g_A^2}{f_\pi^4} \int_{\mid\vec{p}_{1,2}\mid
<k_f}\frac{d^3p_1d^3p_2}{(2\pi)^6}\times} \nonumber \\
{\small \times  \frac{M_B\,(\vec{p_1}-\vec{p_2})^4}{[m_\pi^2+(\vec{p_1}-
\vec{p_2})^2]^2[\Delta^2+\vec{p_2}^2-\vec{p_1}\cdot\vec{p_2}]},} \label{eq:dirquad}
\end{eqnarray}
where we have introduced a suitable change of variables $\vec{p_1}+\vec{l}=\vec{p_2}$.
On the other hand, crossed terms are ignored since their leading order contributions are ${\cal
O}$(1).

\begin{figure*}
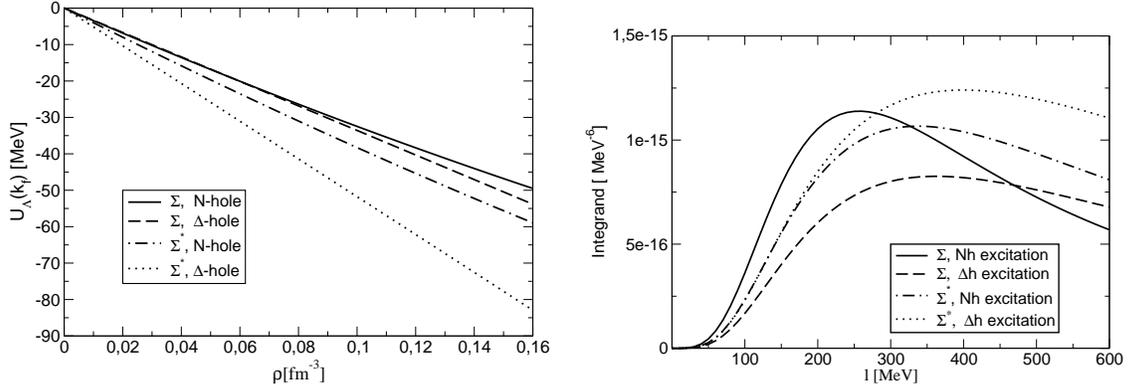

\begin{center}
\begin{tabular}{cc}
\resizebox{0.4\textwidth}{!}{\includegraphics{Pot06.eps}} & \resizebox{0.4\textwidth}{!}{
\includegraphics{Integrand.eps}}
\end{tabular}
\caption{\textbf{Left}: Central potential dependence on nuclear medium density for diagrams presented in Fig.1.  
\textbf{Right}: Integrands of divergent pieces for a particular angular configuration ($\vec{l}\cdot\vec{p_1}=0$) and where a
factor $\vec{l}^2$ has been omitted.} 
\end{center}
\end{figure*}

We compute now the second diagram of Fig.1, which considers a $\Delta$-h instead of a N-h excitation,
integrating the loop energies and expanding to leading order in $M_B$. 
We find that its unique contribution is
\begin{eqnarray}
{\small U_\Lambda(k_f)^{\rm \Delta h} = -\frac{8\,D^2{\cal C}_{\Delta }^2}{9\,f_\pi^4}\int_{\mid\vec{p_1}\mid
<k_f}\frac{d^3p_1d^3l}{(2\pi)^6}\times} \nonumber\\ 
{\small \times \frac{M_B\,\vec{l}^4}{(m_\pi^2+\vec{l}^2)^2(\Delta^2+\Delta_f^2+
\vec{l}^2-\vec{l}\cdot\vec{p_1})}} \label{eq:Delta}.
\end{eqnarray}
The analytical structure of this piece is the same as in (\ref{eq:dirlin}) but with a new splitting term in the
denominator and a different coefficient. The integral  diverges and the previous regulator $\bar{\Lambda}$
is used.

Third and fourth diagrams consider the hyperon $\Sigma^*(1385)$ instead of $\Sigma$ as intermediate
state. A calculation using the procedure shown above leads to similar equations as (\ref{eq:dirlin}), 
(\ref{eq:dirquad}) and (\ref{eq:Delta}) but with different mass splittings in the denominator and different coefficients in 
front of them:
\begin{eqnarray}
{\small U^*_\Lambda(k_f)^{\rm Nh-l} = -\frac{{\cal C}_{\Sigma *}^2 g_A^2}{2\,f_\pi^4}\int_{\mid\vec{p_1}\mid
<k_f}\frac{d^3p_1d^3l}{(2\pi)^6}\times} \\
{\small \times \frac{M_B\,\vec{l}^4}{(m_\pi^2+\vec{l}^2)^2({\Delta^*}^2+
\vec{l}^2-\vec{l}\cdot\vec{p_1})},} \label{eq:Sdirlin} \\
{\small U^*_\Lambda(k_f)^{\rm Nh-q} = \frac{{\cal C}_{\Sigma *}^2 g_A^2}{2\,f_\pi^4} \int_{\mid\vec{p}_{1,2}\mid
<k_f}\frac{d^3p_1d^3p_2}{(2\pi)^6}\times} \nonumber \\
{\small \times  \frac{M_B\,(\vec{p_1}-\vec{p_2})^4}{[m_\pi^2+(\vec{p_1}-
\vec{p_2})^2]^2[{\Delta^*}^2+\vec{p_2}^2-\vec{p_1}\cdot\vec{p_2}]},}
\label{eq:Sdirquad}\\
{\small U^*_\Lambda(k_f)^{\rm \Delta h} = -\frac{{4\,\cal C}_{\Sigma *}^2{\cal C}_{\Delta }^2}{9\,f_\pi^4}\int_{\mid\vec{p_1}\mid
<k_f}\frac{d^3p_1d^3l}{(2\pi)^6}\times} \nonumber\\ 
{\small \times \frac{M_B\,\vec{l}^4}{(m_\pi^2+\vec{l}^2)^2({\Delta^*}^2+\Delta_f^2+
\vec{l}^2-\vec{l}\cdot\vec{p_1})}.} \label{eq:SDelta} 
\end{eqnarray}

Again, the integral of Eq. (\ref{eq:Sdirquad}) is convergent, while the integration of (\ref{eq:Sdirlin}), (\ref{eq:SDelta}) are regularized
with the momentum cut-off $\bar{\Lambda}$.
  
A simple estimation of the amplitudes being integrated can give a hint of the relative importance of the 
different contributions of Fig.1 to the low-momenta interaction. At first glance, decuplet-baryon diagrams are suppressed by
heavier terms in the denominator, but they are also enhanced, on the other hand, by the stronger couplings introduced
by  the $MBB^*$ vertices. That would give, from $\frac{1}{\Delta^2+l^2}$ to
$\frac{2.0}{\Delta^2+\Delta^2_f+l^2}$, $\frac{3.1}{{\Delta^*}^2+l^2}$ or $\frac{6.3}{{\Delta^*}^2+\Delta^2_f+l^2}$
for the comparison between the first diagram (Fig.1) and the second, the third or the fourth respectively. This comparison is
illustrated also in Fig.2 ({\bf Right}) where the integrands of the dominant pieces for a particular angular configuration are drawn
along the interval of integration. All this shows how the
extra damping in the new diagrams introduced by the larger mass splittings is compensated by the stronger couplings of the 
pion to the decuplet. 

The density dependence of the different pieces of the potential is presented in Fig.2 ({\bf Left})). 
The value given by these diagrams at saturation density is strongly attractive
\begin{eqnarray}
U^{\rm l-r}_\Lambda(\rho_0)\simeq-250, {\rm MeV}
\end{eqnarray}
and consequently short-range terms have to be introduced in order to get a realistic central-potential 
depth. See, for instance, Ref. \cite{Tolos:2006ny} where the inclusion of short range correlations
in these pieces leads to reasonable total potentials.

\section{$\Lambda$-nucleus spin-orbit potential}
\label{sec:2}

The spin-orbit coupling is described by the spin-dependent part of the self-energy produced when we
consider the interaction of the corresponding particle (in our case a $\Lambda$-hyperon) with a weakly
inhomogeneous medium. This is achieved \cite{Kaiser:2004fe} considering that the $\Lambda$-hyperon scatters from an initial
trimomentum $\vec{p}-\vec{q}/2$ to a final trimomentum $\vec{p}+\vec{q}/2$. Then, the dominant spin-dependent part
for such weak inhomogeneity arises as
\begin{eqnarray}
\Sigma_{\rm spin}=\frac{i}{2}\vec{\sigma}\cdot (\vec{q}\times\vec{p}) U_{\Lambda ls}(k_{f0}). \label{eq:spinS-E}
\end{eqnarray}

The spin-orbit strength is identified with the shell-model spin-orbit potential when we multiply
by a normalized density profile $f(r)$ and express the result in coordinate space by a Fourier transform:
\begin{eqnarray}
{\cal H}_{\Lambda ls}=U_{\Lambda ls}(k_{f0})\frac{1}{2r}\frac{df(r)}{dr}\,\vec{\sigma}\cdot{\vec{L}},
\label{eq:SOshell}
\end{eqnarray}
where $\vec{L}$ is the orbital angular momentum of the $\Lambda$-hyperon.

The antisymmetric vectorial structure of (\ref{eq:spinS-E}) can be obtained manipulating the expression 
$\vec{\sigma}\cdot (\vec{l}+\vec{q}/2) \vec{\sigma}\cdot (\vec{l}+\vec{q}/2)$ coming  from the $\pi\Sigma
\Lambda$ vertices in the two first diagrams of Fig.1, and also $\vec{S}\cdot (\vec{l}+\vec{q}/2) \vec
{S}^\dagger\cdot (\vec{l}+\vec{q}/2)$  from the $\pi\Sigma^*\Lambda$ vertices of the two last diagrams 
of Fig.1. Using the known relations
\begin{eqnarray}
\sigma_i\sigma_j=\delta_{ij}+i\epsilon_{ijk}\sigma_k \label{octet-spinorial-st}
\end{eqnarray}
and
\begin{eqnarray}
S_iS^\dagger_j=2/3\,\delta_{ij}-i/3\,\epsilon_{ijk}\sigma_k \label{decuplet-spinorial-st}
\end{eqnarray}
we obtain the antisymmetric tensorial structure
which characterizes the spin-dependent part of the self-energy of Eq. (\ref{eq:spinS-E}). Notice that these terms have
a different sign depending on the SU(3)-multiplet which the internal-line baryon belongs to, circumstance that
produces cancellations between the diagrams with $\Sigma$ and the diagrams with $\Sigma^*$. The
other factor $\vec{p}$\, is  in the denominator and comes up after expanding the amplitude in a power series and
keeping the linear term. Finally, 
we obtain the following $\Lambda$-nucleus spin-orbit potentials for the different diagrams:
\begin{eqnarray}
{\small U_{\Lambda ls}(k_f)^{\rm Nh-l} = -\frac{2 D^2 g_A^2}{3 f_\pi^4}\int_{\mid\vec{p_1}\mid
<k_f}\frac{d^3p_1d^3l}{(2\pi)^6}\times} \nonumber \\ 
{\small \times\frac{M_B\,\vec{l}^4}{(m_\pi^2+\vec{l}^2)^2(\Delta^2+
\vec{l}^2-\vec{l}\cdot\vec{p_1})^2}, } \label{eq:SOdirlin}\\
{\small U_{\Lambda ls}(k_f)^{\rm Nh-q} = \frac{2 D^2 g_A^2}{3f_\pi^4} \int_{\mid\vec{p}_{1,2}\mid
<k_f}\frac{d^3p_1d^3p_2}{(2\pi)^6}\times} \nonumber \\
{\small \times  \frac{M_B\,(\vec{p_1}-\vec{p_2})^4}{[m_\pi^2+(\vec{p_1}-
\vec{p_2})^2]^2[\Delta^2+\vec{p_2}^2-\vec{p_1}\cdot\vec{p_2}]^2},} \label{eq:SOdirquad}\\
{\small U_{\Lambda ls}(k_f)^{\rm \Delta h} = -\frac{16\,D^2 {\cal C}_{\Delta }^2}{27\,f_\pi^4}\int_{\mid\vec{p_1}\mid
<k_f}\frac{d^3p_1d^3l}{(2\pi)^6}\times} \nonumber\\ 
{\small\times \frac{M_B\,\vec{l}^4}{(m_\pi^2+\vec{l}^2)^2(\Delta^2+\Delta_f^2+
\vec{l}^2-\vec{l}\cdot\vec{p_1})^2},} \label{eq:SODelta}\\ 
{\small U^*_{\Lambda ls}(k_f)^{\rm Nh-l} = \frac{\,{\cal C}_{\Sigma *}^2 g_A^2}{6\,f_\pi^4}\int_{\mid\vec{p_1}\mid
<k_f}\frac{d^3p_1d^3l}{(2\pi)^6}\times}\nonumber\\
{\small \times\frac{M_B\,\vec{l}^4}{(m_\pi^2+\vec{l}^2)^2({\Delta^*}^2+
\vec{l}^2-\vec{l}\cdot\vec{p_1})^2},} \label{eq:SOSdirlin} \\
{\small U^*_{\Lambda ls}(k_f)^{\rm Nh-q} = -\frac{{\cal C}_{\Sigma *}^2 g_A^2}{6\,f_\pi^4} \int_{\mid\vec{p}_{1,2}\mid
<k_f}\frac{d^3p_1d^3p_2}{(2\pi)^6}\times} \nonumber \\
{\small\times  \frac{M_B\,(\vec{p_1}-\vec{p_2})^4}{[m_\pi^2+(\vec{p_1}-
\vec{p_2})^2]^2[{\Delta^*}^2+\vec{p_2}^2-\vec{p_1}\cdot\vec{p_2}]^2},}
\label{eq:SOSdirquad}\\
{\small U^*_{\Lambda ls}(k_f)^{\rm \Delta h} = \frac{4\,{\cal C}_{\Sigma *}^2{\cal C}_{\Delta }^2}{27\,f_\pi^4}\int_{\mid\vec{p_1}\mid
<k_f}\frac{d^3p_1d^3l}{(2\pi)^6}\times} \nonumber\\ 
{\small \times \frac{M_B\,\vec{l}^4}{(m_\pi^2+\vec{l}^2)^2({\Delta^*}^2+\Delta_f^2+
 \vec{l}^2-\vec{l}\cdot\vec{p_1})^2}.} \label{eq:SOSDelta} 
\end{eqnarray}
All these integrations are convergent and therefore don't depend in other input
parameters than the coupling constants  and particle masses. Notice also that these contributions account 
for non-relativistic effects since they arise at leading 
order in $M_B$ expansion. This is a different situation to that which emerges in mean-field models with 
OBE interactions, where the spin-orbit interaction is a relativistic effect\cite{Brockmann:1977es2}.We have checked
numerically that the expansion in the $M_B$ is good, even when  the mass splittings are almost 300
MeV. The difference at $\rho=\rho_0$ is less than 10\% for all diagrams. 
In Fig. 3,  it is shown the density dependence of the spin-orbit
potentials calculated in this manner.  The $\Delta$-hole diagram (second diagram of Fig.1) gives
a contribution similar in size and of the same sign as the $N$-hole diagram (first). This would spoil the result of 
\cite{Kaiser:2004fe} and produce a too
large negative contribution. However, the processes with a $\Sigma^*$ have a
positive contribution giving a total result quite similar to that obtained
previously including only the first diagram. As explained before, this
different sign comes from the opposite sign of the antisymmetric parts of Eqs.
(\ref{octet-spinorial-st})  and (\ref{decuplet-spinorial-st}), which correspond to octet-octet and
octet-decuplet spin transition operators respectively. 

We also show in Fig. 3 a rough estimate of the
total result by using the same approach as in Ref.  \cite{Kaiser:2004fe} to account for
the missing short range pieces. A full discussion justifying this approach can
be found there. We take

\begin{figure}
\resizebox{0.4\textwidth}{!}{
\includegraphics{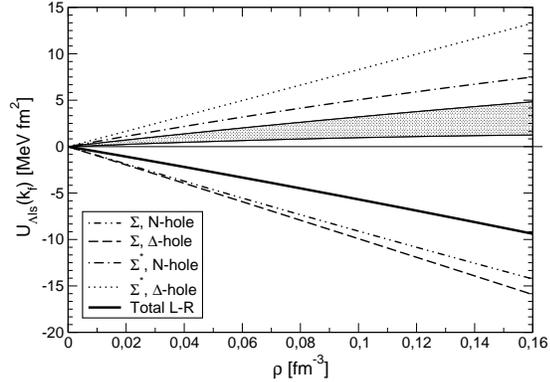}}
\caption{Spin-Orbit potential $U_{\Lambda ls}(k_{f})$ of a $\Lambda$-hyperon in isospin-symmetric nuclear
matter for the Fig.1 diagrams and for the full model added to  the short-range parametrization (shaded area). The short-range
parametrization is $U_{\Lambda ls}^{\rm shell}(k_{f})$=21.3$C_l$\,MeV\,fm$^2$\,$\rho$/$\rho_0$ where $C_l$ lies
between 1/2 and 2/3.}
\end{figure}
\begin{eqnarray}
U_{\Lambda ls}^{\rm shell}(k_{f})=C_l\frac{M_N^2}{M_\Lambda^2}U_{N ls}^{\rm shell}(k_{f})  \label{eq:closerange}
\end{eqnarray}
where the factor $M_N^2/M_\Lambda^2$ comes from the replacement of the nucleon by the $\Lambda$-hyperon in
these relativistic spin-orbit terms. For $U_{N ls}^{\rm shell}(k_{f})$ we suppose a linear dependence in $\rho$
that takes the value 30 MeV\,fm$^3$ at saturation density \cite{Chabanat:1997un}. For  $C_l$ we take the band
between the values 1/2 and  2/3.   

%
%

%
%

\end{document}